# Off-Site Continuums as Cosmological Models of Galaxies and the Universe

© A.V. Novikov-Borodin

Institute for Nuclear research of RAS, Moscow, Russia

Email: novikov@inr.ru; Internet: www.inr.ru/~novikov

**Abstract:** The off-site continuums are proposed for the cosmological models of the galaxies and the Universe. It is shown that many visual properties of galaxies and the Universe may be described on frames of the off-site continuums methodology. In cosmological scale, the appearance of off-site objects is quite similar to the influence of the 'dark matter' and the 'dark energy'. Analogies to known relic radiation also look through. It is discussed few possible models of galaxies and the Universe. Such point of view may appear useful for the investigation of conceptual problems of modern cosmology.

## Introduction: the root of the problem

The general theory of relativity (GR) generates conservation laws inside itself (E. Schrödinger [1]), because four identical relations between Hamiltonian derivatives of some invariant density $\Re$ may be obtained only from the fact of the general invariance of the Integral:

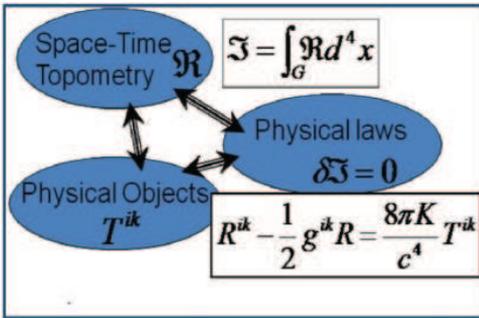

$$\Im = \int_G \Re d^4 x .  \qquad (1)$$

These relations appear like conservation laws and some of such relations have the form of well-known Einstein-Hilbert equations:

$$R^{ik} - \frac{1}{2} g^{ik} R = \frac{8\pi K}{c^4} T^{ik} , \qquad (2)$$

Figure 1. The structure of continuum in general relativity

where $R = g^{ik} R_{ik} = \Re / \sqrt{-g}$ is a curvature of space-time (as usual, the summation on repeating indexes is meant), $g^{ik}$ is a space-time metric tensor, $T^{ik}$ is the energy-momentum tensor of physical objects, $c$ is a speed of light, $K$ is a gravitational constant. This equation defines the physical laws ($\delta \Im = 0$) in continuum and interconnects physical objects ($T^{ik}$) with the space-time topometry (topology and metrics). Therefore, the notion of continuum in GR in fact is not a simple four-dimensional manifold of real numbers and is understood as a System of visual physical reality consisting of physical objects with space-time topometry and corresponding physical laws (see Fig.1). System components are interconnected by Einstein-Hilbert equations Eq.(2). Physics exactly deals with the description of this continuum: physical laws and objects in it. However, do we have some restrictions to existence of another, off-site continuums differed from $G$? If such continuums may exist, so there are other Worlds, off-site space-time continuums, which differ from 'our' World. This way, how could they be observed? Can we describe them mathematically? Do they have some correspondences with well-known physical objects? Whether do we really need such continuums to make our scientific Worldview consistent?

## 1. Space-time continuums

In general theory of relativity the invariant density $\Re$ from Eq.(1) defines the space-time topometry (topology and metrics) and also depends on physical objects existing in this space-time. L.Landau and E.Lifshitz had described this fact in [2] as: *"It is necessary, strictly speaking, to have a set of infinite number of bodies filling all space, like some 'medium'. Such system of bodies together with connected to each of them arbitrarily clocks is a frame of reference in the general theory of relativity"*. Thus, the continuum $G$ with invariant density $\Re$ defines or generates the 'medium', i.e. the system including 'a set of infinite number of bodies' and corresponding conservation laws for these physical objects. We do not have reasons to deny an existence of some OSCs and to reject their possibility to generate its own physical system including its own physical objects, own space-time structure and conservation laws. This 'medium' may differ from one generated by $G$. We have introduced continuums $\widetilde{G}$ differed from the continuum $G$ of the observer as *Off-site* ones (OSCs). If some correspondence $F : G \xleftrightarrow{F} \widetilde{G}$ exists between elements of two continuums, so, generally, it means that there are some regions or subsets $D \subset G$ and $\widetilde{D} \subset \widetilde{G} : G \supset D \xleftrightarrow{F} \widetilde{D} \subset \widetilde{G}$ (see Fig.2). We will call subsets $D$ and $\widetilde{D}$ as *common regions* of continuums $G$ and $\widetilde{G}$. Further in text, we will usually consider continuums from the observer's continuum $G$ point of view. If $\widetilde{D}$ coincides with entire continuum $\widetilde{G}$, so $\widetilde{G}$ will be called as *enclosed* in relation to *containing* $G$. Vice versa, if $D$ coincides with entire continuum $G$, so $G$ will be called as enclosed in relation to containing $\widetilde{G}$.

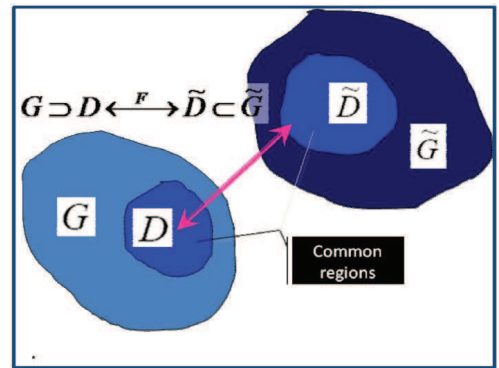

Figure 2. Correspondence between OSCs

Correspondence $F : G \supset D \xleftrightarrow{F} \widetilde{D} \subset \widetilde{G}$ assigned between elements of two multi-dimensional manifolds $G : \left\{ x : (x^0, ..., x^n) \right\}$ and $\widetilde{G} : \left\{ \widetilde{x} : (\widetilde{x}^0, ..., \widetilde{x}^m) \right\}$ may be noted as:

$$\widetilde{x}^j = f^j(x^0, ..., x^n), \ j = 0..m \quad \text{or} \quad x^i = \widetilde{f}^i(\widetilde{x}^0, ..., \widetilde{x}^m), \ i = 0..n. \quad (3)$$

Here, functions $f$ and $\widetilde{f}$ may be of any kind with the fields of definition $D$ and $\widetilde{D}$ correspondingly.

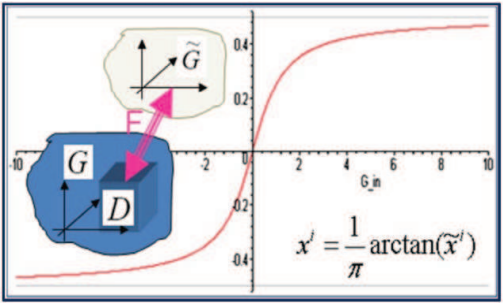

Figure 3. An example of correspondences between manifold

For example, transformations of coordinates realized by any monotone and bounded functions, say $x^i = (1/\pi)\arctan(\widetilde{x}^i)$, $i = 0..n-1$, between two $n$-dimensional manifolds $G : \left\{ x : (x^0, x^1, ..., x^{n-1}) \right\}$ and $\widetilde{G} : \left\{ \widetilde{x} : (\widetilde{x}^0, \widetilde{x}^1, ..., \widetilde{x}^{n-1}) \right\}$ reflect $\widetilde{G}$ inside $n$-dimensional unit cube from $G : \widetilde{G} \leftrightarrow D_1 \subset G$ (see Fig.3). On Figure 3 $\widetilde{G}$ is enclosed in relation to $G$. Another transformations $\widetilde{x}^i = (1/\pi)\arctan(x^i)$ will reflect $G$ inside $n$-dimensional unit cube from: $G \leftrightarrow \widetilde{D}_1 \subset \widetilde{G}$.

If we suppose that the off-site continuum $\widetilde{G}$ has its own invariant density $\widetilde{\Re}$, so we may write analogously to Eq.(1) the off-site Integral: $\widetilde{\Im} = \int_{\widetilde{G}} \widetilde{\Re} d^m \widetilde{x}$. If it is supposed that $\widetilde{G}$ has some metrics, so we may consider that some interval $d\widetilde{s}^2 = \widetilde{g}_{jl} d\widetilde{x}^j d\widetilde{x}^l$, where $j, l = 0..m$, is introduced. Note that from OSC definition $\widetilde{\Re}$ and $d\widetilde{s}^2$ have to differ from $\Re$ and $ds^2$, which, for example, is not true for coordinate transformations between inertial frames.

Not only the OSC topometry (topology and metrics), but also the transformation functions $f$ and $\widetilde{f}$ from Eq.(3) are very important for observability of continuums. For example, if each element of two identical one-dimensional manifolds $G : x \in (-\infty, \infty)$ and $\widetilde{G} : \widetilde{x} \in (-\infty, \infty)$ is represented as a decimal number: $x = ...a_1, a_0, a_{-1}, a_{-2}...$; $\widetilde{x} = ...b_1, b_0, b_{-1}, b_{-2}...$, where $a_i, b_j = 0,1..9$ and the correspondence $G \leftrightarrow \widetilde{G} : b_{2k+1} = a_{2k}, b_{2k} = a_{2k+1}$, $k = 0, \pm 1, \pm 2, ...$ is introduced, so, in spite of both $G$ and $\widetilde{G}$ are continuous and measurable, and the correspondence is biunique, the 'mutual visible metrics' can't be introduced. We understand as 'mutual visible metrics' the metrics in one continuum observable or visible from another one.

Some particular case of the OSC methodology is an introduction in many modern physical theories of some additional hidden or compact spaces. From OSC's point of view, it means that the OSC dimension is more than the dimension of the observer's continuum: $m > n$ and $n$ from $m$ components of the OSC is identical to the corresponding ones from the observer's continuum. It is considered that the rest $m - n$ components are responsible for some 'internal' parameters of the OSC object. Of course, it is only one particular case, quite specific type of transformations between OSC continuums. Such approaches are based on the ungrounded supposition of uniqueness of 'our' continuum, the continuum of the observer. Exactly this uniqueness is considered as controversial in this paper.

Mathematical and physical correspondences between continuums usually do not coincide. For example, coordinate transformations for uniformly rotating with some frequency $\omega$ frame of references are 'mathematically' biunique for a whole 4D continuum, but 'physically' it may be used up to distance $r = c/\omega$ from the axis of rotation ($c$ is a speed of light), because, as it is usually declared: "such system may not be realized by real bodies" [2].

If functions $f$ and $\widetilde{f}$ from Eq.(3) are continuously differentiable, so, on the field of definition, using $d\widetilde{x}^j = (\partial f^j / \partial x^i) dx^i = (\partial \widetilde{x}^j / \partial x^i) dx^i$, one can get the expression for the visible off-site metrics ($d\widetilde{s}^2 \leftrightarrow ds^2$):

$$d\widetilde{s}^2 = \widetilde{g}_{jl}|_{\widetilde{x} = f(x)} \frac{\partial \widetilde{x}^j}{\partial x^i} \frac{\partial \widetilde{x}^l}{\partial x^k} dx^i dx^k \overset{def}{=} \widetilde{\widetilde{g}}_{ik} dx^i dx^k; \ i,k = 0..n; \ j,l = 0..m. \quad (4)$$

Generally, the visual metric tensor $\widetilde{\widetilde{g}}_{ik} = \widetilde{g}_{jl|\widetilde{x}=f(x)} \dfrac{\partial \widetilde{x}^{\,j}}{\partial x^i} \dfrac{\partial \widetilde{x}^{\,l}}{\partial x^k}$ is not a tensor in $G$ and is defined only in common regions. Only in common regions it is possible to observe the visible part of the OSC Integral, so:

$$\widetilde{\mathfrak{I}} = \int_{\widetilde{G}} \widetilde{\mathfrak{R}} d^m \widetilde{x} \mapsto \int_D \widetilde{\mathfrak{R}} d^m \widetilde{x} = \int_D \widetilde{\mathfrak{R}}_{|\widetilde{x}=f(x)} \left| \dfrac{\partial \widetilde{x}^{\,j}}{\partial x^i} \right| d^n x \overset{def}{\equiv} \int_D \widetilde{\widetilde{\mathfrak{R}}}(x) d^n x , \qquad (5)$$

where $\left| \partial \widetilde{x}^{\,j} / \partial x^i \right|$ is a functional determinant. The functional determinant needs to be defined, so Eq.(5) may be used with $n = m$, which, generally speaking, is not necessary in Eq.(4). To overcome such limitations some 'mathematical tricks' like the considered example with hidden spaces are usually used. Above correspondences Eq.(5) may be represented by equalities in important particular cases of the enclosed ($\widetilde{G}_e$: $\widetilde{G}_e \leftrightarrow D \subset G$) and containing ($\widetilde{G}_c$: $G \leftrightarrow \widetilde{D}_c \subset \widetilde{G}_c$) OSCs:

$$\widetilde{\mathfrak{I}}_e = \int_{\widetilde{G}_e} \widetilde{\mathfrak{R}}_e d^m \widetilde{x}_e = \int_D \widetilde{\mathfrak{R}}_{e|\widetilde{x}=f(x)} \left| \dfrac{\partial \widetilde{x}_e^{\,j}}{\partial x^i} \right| d^n x \text{ , (6a) } \quad \mathfrak{I} = \int_G \mathfrak{R} d^n x = \int_{\widetilde{D}_c} \mathfrak{R}_{|\widetilde{x}=f(x)} \left| \dfrac{\partial x^i}{\partial \widetilde{x}_c^{\,j}} \right| d^m \widetilde{x}_c \text{ . (6b)}$$

## 2. Visible topology and metrics of Off-site continuums

Before starting the mathematical analysis of the off-site continuums, it is necessary to make clear the physical background of the proposed idea of OSC existence. Further in paper we will analyze the off-site continuums from 'our' four-dimensional continuum $G : \left\{ x : \left( x^0, x^1, ..., x^n \right) \right\}$, $n = 3$ and will denote the time as $dt = d\tau / c = dx^0 / c$. To investigate the visible structure or topometry of continuums, we need to suppose that, at least, OSCs have such topometry by themselves. So, we consider that some metrics $d\widetilde{s}^{\,2} = \widetilde{g}_{jl} d\widetilde{x}^{\,j} d\widetilde{x}^{\,l}$, $j,l = 0..m$ is defined in $\widetilde{G}$. In this section the kind of transformations between OSC is not important to us. We will analyze the correspondences between metric tensor $g_{ik}$ of the observer's continuum and the perception $\widetilde{\widetilde{g}}_{ik}$ of OSC metric tensor $\widetilde{g}_{jl}$ (Eq.(4)).

L. Landau and E. Lifshitz have given the following physical interpretation of parameters of 'usual' metric tensor in continuum of the observer [2]: *"It is necessary to emphasize a difference between meanings of a condition $g_{00} > 0$ and a condition of a certain signature (signs on principal values) of the metric tensor $g_{ik}$. The tensor $g_{ik}$, non-satisfying to the second one of these conditions, cannot correspond to any real gravitational field at all, i.e. the metrics of the real space-time. Non-fulfillment of the condition $g_{00} > 0$ would meant only, that the corresponding frame of references can't be realized by real bodies; thus if the condition on principal values is carried out, it is possible to achieve to $g_{00}$ becomes positive by appropriate transformation of coordinates".*

Generally, the visible OSC metric tensor $\widetilde{\widetilde{g}}_{ik}$ is not at all a tensor in $G$, but we will consider that its parameters determine the observable physical space-time structure of $\widetilde{G}$ and the visual properties of the off-site physical objects. The schematic drawings of possible structures of OSCs are presented on Fig.4. Following by L.Landau and E.Lifshitz [2] we may conditionally separate the visual structure of the off-site continuum into three regions: 1) *the timelike region* ($\widetilde{\widetilde{g}}_{00} > 0$ and $\det(\widetilde{\widetilde{g}}_{ik}) < 0$); 2) *the spacelike region* ($\det(\widetilde{\widetilde{g}}_{ik}) > 0$) and 3) *the transitive region* covering the rest part ($\widetilde{\widetilde{g}}_{00} < 0$ and $\det(\widetilde{\widetilde{g}}_{ik}) < 0$). Here, we've replaced a condition of the certain signature of the metric tensor by the condition of the sign of $\det(\widetilde{\widetilde{g}}_{ik})$, because, generally, $\widetilde{\widetilde{g}}_{ik}$ is not a tensor in $G$. It is also shown the fourth principally invisible *non-observable region* in OSC, which, of course, even could not be imagined in [2], because it is 'outside' the observer's continuum. It is a rest part of OSC without common regions.

It is shown on Fig.4 three possible variants of correspondences of the OSC $\widetilde{G}$ and the continuum of the observer $G$:

A. $G / D \neq \emptyset$ and $\widetilde{G} / \widetilde{D} \neq \emptyset$;

B. $\widetilde{G}$ is completely included in $G$ ($\widetilde{G}$ is an enclosed continuum); and

C. $\widetilde{G}$ completely contains $G$ ($\widetilde{G}$ is a containing continuum).

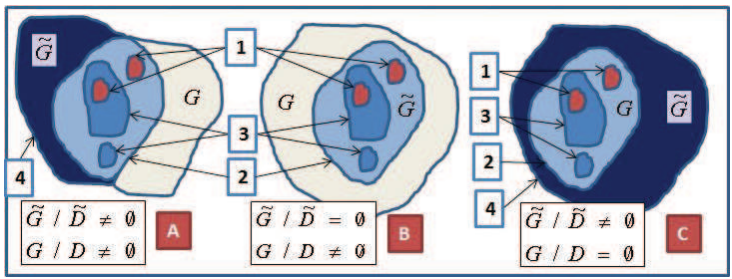

Figure 4. The schematic drawings of the visible OSC regions:
1) timelike; 2) spacelike; 3) transitive; 4) invisible.

If $\widetilde{G}$ has non-common regions ($\widetilde{G}/\widetilde{D} \neq \emptyset$), these regions will be principally non-observable from $G$. These regions (region 4) exist in A and C variants. Generally speaking, there exists another variant, where $G$ and $\widetilde{G}$ have no any common regions at all, but for the OSC to be observed it needs to be some 'media' between continuums, so this variant may be considered as a particular case of variants A or C.

**Timelike regions** ($\widetilde{\widetilde{g}}_{00} > 0$ and $\det(\widetilde{\widetilde{g}}_{ik}) < 0$). These are regions of 'normal' matter. The visible time and distance in OSC may be defined, following by L.Landau and E.Lifshitz [2], as: $d\widetilde{\widetilde{\tau}} = \sqrt{\widetilde{\widetilde{g}}_{00}}\, dx^0$ and $d\widetilde{\widetilde{l}}\,^2 = \widetilde{\widetilde{\gamma}}_{\alpha\beta} dx^\alpha dx^\beta$, $\widetilde{\widetilde{\gamma}}_{\alpha\beta} = -\widetilde{\widetilde{g}}_{\alpha\beta} + \widetilde{\widetilde{g}}_{0\beta}\widetilde{\widetilde{g}}_{0\beta}/\widetilde{\widetilde{g}}_{00}$. Hence, the physical correspondence may be introduced and OSC objects may be identified as visual 'normal matter' from $G$. As far as the motion of OSC objects in timelike regions is defined by their own topometry, this motion will be observed from $G$ as quite unusual. For the observer it will look like the action of some 'forces' applied to these OSC objects, holding them inside some region from $D_\alpha$ ($D : D_\alpha \times \tau$, so $D_\alpha$ is understood as a cross section of common regions $D$ at each $\tau$). Exactly such 'forces' were introduced in quantum physics to explain the capturing of charged particles inside the nuclei. Analogies with strong interactions, phenomenon of confinement are arising at once. It will be shown in next section that such 'forces' really have obvious correlations with strong interactions and string models. We do not have any reason to decline such processes in cosmological scales, but these processes will have some specifics: $D_\alpha$ may not be limited in space, also it is not clear how identify the boundaries of $D_\alpha$, which may also be time-dependent.

**Transitive regions** ($\widetilde{\widetilde{g}}_{00} < 0$ and $\det(\widetilde{\widetilde{g}}_{ik}) < 0$). "Non-fulfillment of the condition $g_{00} > 0$ would mean only, that the corresponding frame of references can't be realized by real bodies; thus if the condition on principal values is carried out, it is possible to achieve to $g_{00}$ becomes positive by appropriate transformation of coordinates" [2]. Therefore, OSC objects may have different perception in transitive regions depending on concrete reference frame in the observer's continuum. As far as physical space correspondences are not able to introduce, OSC physical objects from transitive regions will be observed as something amorphous, distributed in space. However, from another reference frame in $G$, these OSC physical objects may look like 'real bodies'. Therefore, the OSC objects from transitive regions will look like strange 'amorphous' media distributed in some space regions, but possessing some 'real' physical characteristics, may interact as 'real body'. For example, real massive body identified in one reference frame may not be identified as this body from another frame, but its gravitational action to other bodies would not disappear for the observer. Even from such schematic view, one may notice the correlations of transitive OSC objects with *dark matter*. Indeed, dark matter is understood as some invisible distributed in space substance, strange 'amorphous' media interacting gravitationally with identified visible objects.

**Spacelike regions** ($\det(\widetilde{\widetilde{g}}_{ik}) > 0$). The observation of OSC physical objects from spacelike regions may be even more unusual than the OSC from transitive regions: "The tensor $g_{ik}$ cannot correspond to any real gravitational field at all, i.e. the metrics of the real space-time" [2]. The spacelike OSC physical objects also will be observed as some amorphous media, distributed in space, but also the macro-characteristics of these objects couldn't be identified with corresponding characteristics of 'normal' matter, 'real bodies', 'real gravitational field'. Therefore, the OSC physical objects in spacelike regions may possess quite unusual, even unphysical characteristics. Such unusual characteristic as negative pressure is used for the description of *dark energy* in Einstein's cosmological principle (Y. Baryshev [3]).

**Non-observable regions.** It is impossible to observe the OSC physical objects from the non-observable regions, but it does not mean, that it is impossible to register their influence or action on the physical objects from the continuum of the observer. It might be streams of particles flowing somewhere from the invisible source or disappearing somewhere in space. This way, if physical 'matter exchange' with non-observable regions really exist, it may lead to visual violations of conservation laws. Note that similar violation of the energy conservation law is a real problem in the standard model of the expanding Universe [3]. The physical objects from non-observable regions influence to the metrics on common regions, so also may be interpreted as influence of 'dark energy', although the 'dark energy' from spacelike regions has another nature. Note, that the observer is a part of his own continuum, so he may 'physically' use only invariant transformations for observations, so such separation is fixed for him and only in transitive regions the multiple perception is possible.

Such 'strange' properties of the physical objects from different OSC regions may only be their perceptions. Without knowing the transformation laws, we cannot say anything about their 'real' properties in 'home' continuum. So, may be, even 'visual forces' observed between OSC objects do not exist in their 'home' OSC. Also note, that the visible OSC metric tensor usually is time-dependent in the system of the observer, so the OSC internal structure is perceived dynamical. The time-scale of the observation may be quite different for different OSCs.

Thus, we have supposed the physical existence of the OSCs together with their 'own' objects and have found some analogies of these objects with known physical objects. But it, for example, means that one can observe the 'real' or 'normal' matter, dark matter and even dark energy 'locally' in space (variants A and B on Fig.4) or 'globally' (variant C). Last data from the Hubble space telescope show that such localized objects really exist. On Fig.5A, the three-dimensional map offers a first look at the web-like large-scale distribution of dark matter, an invisible form of matter that accounts for most of the universe's mass. The image shows that the supercluster galaxies lie within the clumps of dark matter (Credit: NASA, ESA, R. Massey (Caltech)).

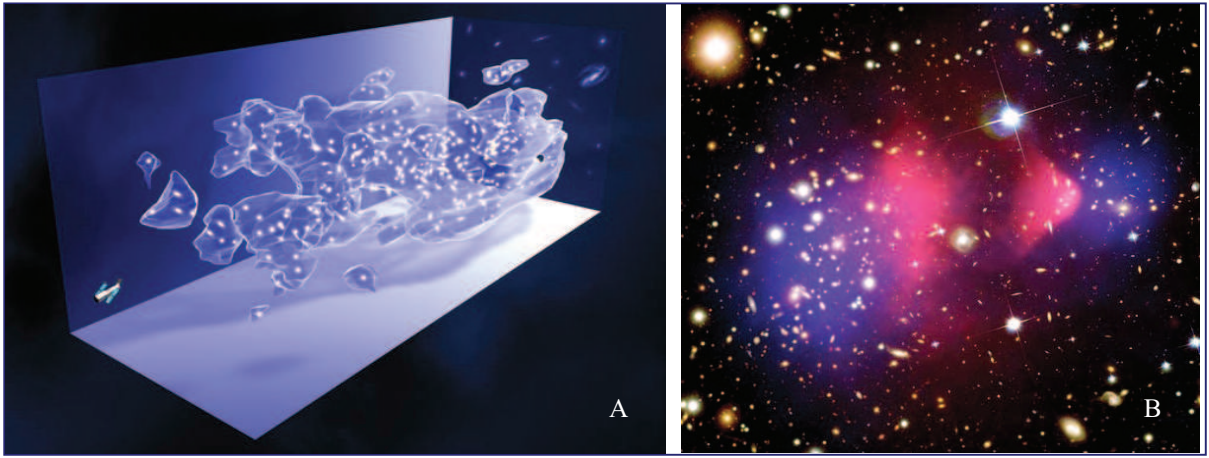

Figure 5. The dark and normal matter distributions: A) the 3D map of the large-scale distribution of dark matter; B) the collision of two large clusters of galaxies 1E 0657-56.

There are some reports about the HST observations of the local dark energy, which also confirms our OSC model. The group of authors in papers [4,5] estimated the local dark energy nearby galaxy groups M81/82 and Cen A/M83. They conclude that the local density of dark energy in the area is to be near the global dark energy density or perhaps exactly equal to it. This fact also is in a good agreement with the OSC model.

It is interesting to make some 'huge' estimations of 'amount of matter' in different OSC regions (see Fig.6). Indeed, supposing that there is equal possibility to observe $\widetilde{\widetilde{g}}_{00}$ and $\det(\widetilde{\widetilde{g}}_{ik})$ positive or negative, so in enclosed OSC one will observe 25% of matter in timelike ($h_e^1$) and transitive regions ($h_e^2$), and 50% in spacelike region ($h_e^3$). Modern observed data of matter in the Universe are about 5% of 'normal' matter ($h_o^1$), 25% of dark matter ($h_o^2$) and 70% of dark energy ($h_o^3$). It is possible to recalculate the compound of matter in rest OSCs (OSCs with non-observable regions, i.e. containing and others): $h_c^i = (1+\eta)h_o^i - \eta h_e^i$, where $\eta = Me/Mc$ is a ratio between 'amounts of matter' in enclosed ($Me$) and other ($Mc$) continuums. For any continuum $\sum h^i = 1$ and any $h^i > 0$. To satisfy $h_c^1 \geq 0$ it needs to be $\eta = Me/Mc \leq 0.25$.

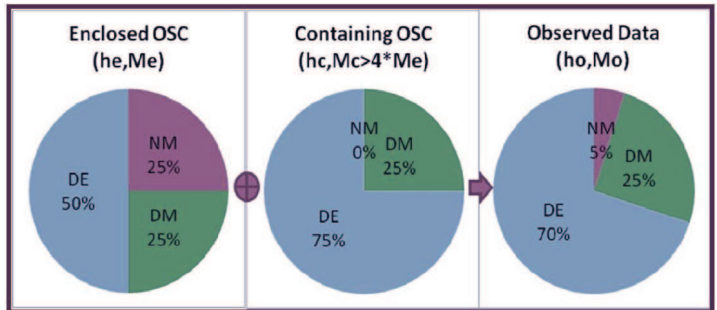

Figure 6. 'Matter' in OSC regions

The 'geometrical' ratio between dark and normal matter may be seen on Fig.5B at the tremendous collision of two large clusters of galaxies. Dark matter and normal matter have been wrenched apart. This composite image shows the galaxy cluster 1E 0657-56, also known as the 'bullet cluster'. The hot gas detected by 'Chandra' in X-rays is seen as two pink clumps in the image and contains most of the 'normal' matter in the two clusters. The bullet-shaped clump on the right is the hot gas from one cluster, which passed through the hot gas from the other larger cluster during the collision. The geometrical areas of the normal and dark matter on Fig.5B approximately equal to each other, which corresponds to our huge estimations. Last observations of the Universe give 75% for dark energy, 25% for dark matter and only 5% for 'normal' matter. Hence, the 'global' dark energy and dark matter in observable parts of containing and other OSCs prevail in 'our' continuum. Everyone can ask how normal is 'normal' matter?

## 3. Physical Description

Let's consider the *'global system'* containing *all* interconnected continuums (and nothing else), so this system is *closed* (closed-loop) and *isolated* by definition. As any system, the global one needs to be consistent, non-contradictive (in contrary it would not be a system), needs to have some 'global laws' unifying different OSCs as subsystems inside one system. Physical laws of any OSC, including our continuum, cannot contradict to physical laws of the 'global' system. Hence, the global physical equations need to coincide with general description of any OSC in limit, i.e. if only one continuum is considered. All above conditions may be satisfied if to consider the corresponded 'Integral' of global system analogously to the 'general' Integral Eq.(1) of our continuum:

$$\mathfrak{I}_S = \int_{\cup \widetilde{G}} \Lambda\left(\widetilde{\mathfrak{R}}_{(0)}, \widetilde{\mathfrak{R}}_{(1)},...,\widetilde{\mathfrak{R}}_{(n)},...\right) d\Omega^* \equiv \int_{G_S} \Lambda\left(\widetilde{\mathfrak{R}}\right) d\Omega^* . \qquad (7)$$

Here $d\Omega^*$ is some generalized elementary volume of 'combined' continuums $G_S = \cup \widetilde{G}_{(i)}$ and $\widetilde{\mathfrak{R}}$ is a set of scalar densities of all possible OSCs. To simplify notations it is applied $\mathfrak{R} = \widetilde{\mathfrak{R}}_{(0)}$ and $G = \widetilde{G}_{(0)}$. Generally, the elementary volume $d\Omega^*$ cannot be expressed on common variables, because even the fields of definition of scalar densities of different OSCs are not coincide (see Fig.2). Taking into account Eq.(3) one can express $d\Omega^*_{(n)}$ on common regions in observer's continuum as $d\Omega^*_{(n)} = \left|\partial \widetilde{x}^j_{(n)} / \partial x^i\right| d\Omega$, where $\left|\partial \widetilde{x}^j_{(n)} / \partial x^i\right|$ is a functional determinant. This way the resulted part of the OSC scalar density will be $\widetilde{\widetilde{\mathfrak{R}}}_{(n)} = \widetilde{\mathfrak{R}}_{(n)|\widetilde{x}=f(x)}\left|\partial \widetilde{x}^j_{(n)} / \partial x^i\right|$. Of course, if functions Eq.(3) are not differentiable, such equations have no mathematical sense. As far as we do not want to exclude functions of any kind, here and later we will understand such expressions symbolically, in *'extended mathematical sense'*, whenever it needs to.

To make the Global system consistent, the Integral $\mathfrak{I}_S$ Eq.(7) needs to coincide with Eq.(1) if only one observer's continuum is considered. It needs to be true for any OSC, so one can demand from function $\Lambda(\widetilde{\mathfrak{R}})$: $\left\{\forall n : \Lambda(\widetilde{\mathfrak{R}}_{(n)}) = \widetilde{\mathfrak{R}}_{(n)}\right\}$. Furthermore, if functions $\widetilde{\mathfrak{R}}_{(n)}$ have no common regions, so, are independent, then: $\Lambda(\widetilde{\mathfrak{R}}) = \sum \widetilde{\mathfrak{R}}_{(n)}$. From other side, if fields of definition of OSCs completely coincide, $\Lambda(\widetilde{\mathfrak{R}})$ needs to be the scalar density, so it may be equal to any $\widetilde{\mathfrak{R}}_{(n)}$ or their accidental linear combinations.

This conclusion looks strange and ambiguous from the first sight, but in fact, it already *was* a warning in GR and *is* logical consequence of OSC approach. Indeed, E. Schrodinger had mentioned in [1] that in GR the identical relations gotten from variation of the Integral Eq.(1) *"are not the only ones. Any scalar density creates some system of identities. We would prefer to get some assertion concerned with only one concrete density* $\mathfrak{R}$ *and even not necessary the identity. <...> from other side, the conservation laws by themselves represent the only one separated fact, but not the kind of facts".* These 'mathematical' conclusions of E. Schrodinger are 'physically' confirmed in OSC by declaring that not only one concrete density may be physically realized. Let us notice, that in comparison with GR, in OSC much more various scalar densities may be considered, even with different fields of definition.

Thus, we may consider that *'matter'* may be organized or realized in some OSCs $\widetilde{G}_{(n)}$ characterized by some corresponding scalar density $\widetilde{\mathfrak{R}}_{(n)}$ or by combinations of scalar densities. This way, one may symbolically write the global Integral and general scalar density of the OSC system as:

$$\mathfrak{I}_S = \sum_n \widetilde{\mathfrak{I}}_{(n)} = \int_{G_S} \Lambda\left(\widetilde{\mathfrak{R}}\right) d\Omega^*, \quad \Lambda\left(\widetilde{\mathfrak{R}}\right) = a_0\widetilde{\mathfrak{R}}_{(0)} \oplus a_1\widetilde{\mathfrak{R}}_{(1)} \oplus ... \oplus a_n\widetilde{\mathfrak{R}}_{(n)} \oplus ... . \quad (8)$$

Here the coefficients $a_n$ look like 'measure' between OSCs, characterizing the comparative amount of 'matter' realized in different OSC. These coefficients also reflect the degree of state realization: if some $a_n$ is zero, so the $n$-th state is not physically realized, but there exists a possibility of its realization. This way one may consider an OSC as some physically realized *'state'* of the global system. It is not straight summation of scalar densities in $\Lambda(\widetilde{\mathfrak{R}})$, due to their interconnections on common regions. It reflects the fact of their mutual dependence.

According to OSC methodology, one can get Einstein-Hilbert equations for any OSC. So it is possible to consider the variation of system Integral $\mathfrak{I}_S$ Eq.(8). Applying the 'standard' Einstein-Hilbert method [1,2] of getting gravitational 'field' equations for any OSC, it is possible to get generalized 'field' equations for global system:

$$R^{ik} - \frac{1}{2}g^{ik}R - \frac{8\pi K}{c^4}T^{ik} = -\frac{1}{a_0}\sum_{n,n\neq 0} a_n\left(\widetilde{R}^{jl}_{(n)} - \frac{1}{2}\widetilde{g}^{jl}_{(n)}\widetilde{R}_{(n)} - \frac{8\pi K}{c^4}\widetilde{T}^{jl}_{(n)}\right)\frac{\partial x^i}{\partial \widetilde{x}^j_{(n)}}\frac{\partial x^k}{\partial \widetilde{x}^l_{(n)}} . \quad (9)$$

Here, $\widetilde{R}_{(n)} = \widetilde{g}^{ik}_{(n)}\widetilde{R}_{ik(n)} = \widetilde{\mathfrak{R}}_{(n)}/\sqrt{-\widetilde{g}_{(n)}}$ is the 'internal' curvature and metric tensor in corresponding OSC, $R = g^{ik}R_{ik} = \mathfrak{R}/\sqrt{-g}$ are the curvature and metric tensor of the observer's continuum. Terms $\partial x^i \partial x^k / \partial \widetilde{x}^j_{(n)}\partial \widetilde{x}^l_{(n)}$ are needed to express Eq.(9) in variables of observer's continuum. 'Matter' coefficient $a_0 \neq 0$, because this state is realized at least for the observer. Equations (9) are presented for the observer's continuum, but it possible to get them for any OSC only by changing corresponding indexes. Remember that we understand these expressions and also Eq.(9) in 'extended mathematical sense' mentioned above.

It is possible to give the simple physical interpretation of Eq.(9). Indeed, terms $\widetilde{R}_{(n)}^{jl} - \frac{1}{2}\widetilde{g}_{(n)}^{jl}\widetilde{R}_{(n)}$ include the influence of all $\widetilde{T}_{(n)}^{ik}$ from $\widetilde{G}_n$, while terms $\widetilde{T}_{(n)}^{jl}(\partial x^i \partial x^k / \partial \widetilde{x}_{(n)}^j \partial \widetilde{x}_{(n)}^l)$ correspond to physical objects only from common regions. They need to be rejected, because they are already taken into account in the continuum of the observer by corresponding $T^{ik}$ on terms left. Therefore, terms right are responsible for the influence of the OSC physical objects from non-common regions. In GR the non-common regions are not considered at all, so in Eq.(9) the terms right are applied equal to zero. It is exactly the usual Einstein-Hilbert equation.

Now we need to consider the sense of energy-momentum tensors in Eq.(9) from OSC point of view.

# 4. OSC Physical Objects

Let us consider the possible ways of evolution of some physically realized and *closed* subsystem $\widetilde{G}_{Sub} \subset G_S$. The subsystem may differ from OSC, coincide with it or with few OSCs. It will be important for us that the subsystem is closed, so the 'amount of matter' is unchangeable in it and $\delta(\widetilde{\mathfrak{I}}_{Sub}) = 0$, so $\widetilde{\mathfrak{I}}_{Sub}$ is invariant in $G_S$. Analyzing Eq.(6), one can see that even some closed OSC $\widetilde{G}_n$, i.e. for which there are no matter exchange with global system and $\delta(\widetilde{\mathfrak{I}}_n) = 0$, can't be considered as independent or isolated, because it changes the curvature of global system, so takes part in gravitational interaction. Mathematically it means, that the condition $\delta(\mathfrak{I}_S) = 0$ is not equivalent to $\delta(\widetilde{\mathfrak{I}}_n) = 0$ for any $n$. Of course, this conclusion is also valid for subsystem.

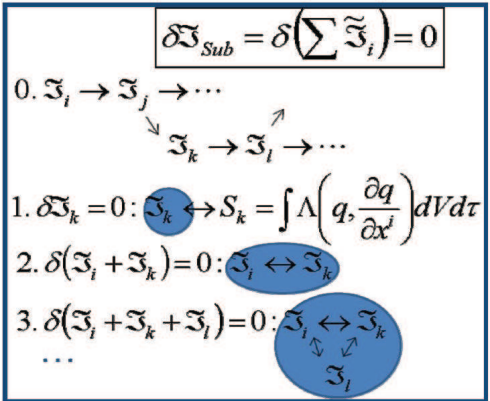

Figure 7. OSC Physical Objects

As far as the subsystem is closed, so matter cannot disappear from it. Disappearing in one state of the subsystem, matter needs to appear in other (or others). There are initiated 'open' chains of excitations of different unstable states. This process is schematically shown on the variant 0 of Fig.7. Unstable states are consequently excited ones by ones. It corresponds, for example, to permanent excitation of unstable (virtual) particles and looks like a model of physical vacuum.

If the closed subsystem coincides with some stable state $\widetilde{G}$ (we missed indexes to simplify notations) of the global system, so some Integral $\widetilde{\mathfrak{I}} = \int_{\widetilde{G}} \widetilde{\mathfrak{R}} d\widetilde{\Omega}$ is defined and it is invariant in global system. The Integral $\widetilde{\mathfrak{I}}$ will be seen invariant only if this stable state corresponds to the enclosed OSC ($\widetilde{D} \equiv \widetilde{G}$ on Fig.2 and non-common regions do not exist for the observer), otherwise the matter exchange between common and non-common regions is possible. Hence, for the enclosed OSC the above Integral in the observer's continuum coincides with the action of some closed subsystem [2]:

$$\widetilde{\mathfrak{I}} = \int_D \widetilde{\mathfrak{R}}|_{\widetilde{x}=f(x)} \left| \frac{\partial \widetilde{x}^j}{\partial x^i} \right| d\Omega \leftrightarrow \int_D \Lambda\left(q, \frac{\partial q}{\partial x^i}\right) \sqrt{-g} \, d\Omega = cS \ . \tag{10}$$

This realization of subsystem as a stable OSC state is shown on the variant 1 of Fig.7. It is possible in GR by general rules to get the energy-momentum tensor of any physical system, which action is represented as Eq.(10) [2]. This way it is possible to get expressions for energy-momentum tensors of macroscopic bodies and the electro-magnetic field.

If the Integral of OSC is invariant, it may be put into the correspondence with the integral $-mc^2 \int ds$, where $ds$ is the interval of some fixed point from $D$. If also the space cross-section $D_\alpha$ of $D$ is limited in $G$, so the principally non-local OSC object will behave 'in a whole' as a free material point with rest mass $m$.

There also exist other possibilities. Excited unstable states may form the 'closed' chain. Many unstable states may be involved. Variants 2 and 3 on Fig.7 illustrate the subsystem with two and three unstable states. Of course, it may be more. Unstable states here are in permanent matter exchange with each other. As far as each of them is unstable, it is impossible to observe them outside of the subsystem. Such stable subsystem of unstable states is also the closed physical system and may be described by Eq.(10), but 'as a whole' with its 'internal structure'. This physical model is very similar to quark-gluon models in chromo dynamics. The string theories (see, for example, L.D. Soloviev [6,7] and

A. Polyakov [8]) are using the action of kind of Eq.(10). For example, L.D. Soloviev in [6] used an action of relativistic quantum model of string with the curvature in the form $S = \int_{\tau_1}^{\tau_2} \int_{\sigma_1}^{\sigma_2} F(R/2)\sqrt{-g}\,d\sigma d\tau$, where $R$ is a curvature on $\sigma$. The string models differ from each other mainly by the concrete form of $\Lambda$.

The OSC metrics $\widetilde{g}^{jl}$ differs considerably from metrics $\widetilde{g}^{jl}(\partial x^i \partial x^k / \partial \widetilde{x}^j \partial \widetilde{x}^l)$ visual by the observer, so OSC objects are not seemed as captured inside common regions due to their own gravitation, as it is, for example, in case of black holes or even star systems etc. They will be observed as captured for some unknown reasons. This effect is described in quantum physics by introducing the new 'fundamental' interaction – the strong force. The nature of such 'fundamental' interaction is quite clear in OSC model: it is visual OSC topology.

## 5. Excited fields of charged OSC physical objects

There is one problem, one 'cloud on the horizon' in the OSC model. OSC objects may be observed in quite nonlinear motion at least in timelike regions or in dynamics in others. However, if some of them have an electric charge, they would excite the electromagnetic (em) fields in the continuum of the observer loosing the energy and, so, organizing the 'matter exchange' between continuums. Hence, OSCs with charged physical objects will be non-stable. However, we know that 'normal matter' described as OSC objects mainly consists of stable charged particles: electrons, protons, nuclei etc. By the way, it is not a trivial problem and not a problem only in OSC approach, but in modern physics too. Remember, this paradox is not explained in quantum physics and the existence of some stable states is simply postulated. It is unacceptable for us, because our aim is to create the consistent system of physical reality.

So, we are going to consider the question of stability of the closed subsystem with charged physical objects and action like Eq.(10). We shall consider the evolution of the isolated subsystem instead of closed, so independent from the 'external' influence of other physical objects from the global system. Anyway, such influence may be quite accidental, so we shall neglect it now, considering idealized case with the metrics, where the subsystem exists, flat, i.e. $g = -1$. We shall also neglect the gravitation caused by the subsystem itself in the observer's continuum, because it is much less than the em forces. Remember, as mentioned before, it does not contradict to visual 'capturing' of OSC objects in common regions.

According to [2] and taking into account Eq.(10) the action of the enclosed OSC with charged 'matter' on common regions $D$ in observer's continuum is:

$$S = -c\int_D \mu\,d\Omega - \frac{1}{c^2}\int_D A_i j^i d\Omega - \frac{1}{16\pi c}\int_D F_{ik} F^{ik} d\Omega, \qquad (11)$$

where $A_k$ and $F_{ik}$ are the 4D potential and tensor of the em fields, $j^i = c\rho\,dx^i / d\tau$, $\mu$ and $\rho$ are understood as some mass and charge distributions on $D : D_\alpha \times \tau$. For the enclosed OSC object to be stable, we shall demand the zero variation of two last terms right[1]. It leads to the *wave equations*. To provide the stability of the OSC system we need to protect it from the 'matter exchange', so we should find such solutions of the wave equation where excited em fields are localized in $D_\alpha$, so also to satisfy to the edge conditions:

$$g^{lm}\frac{\partial^2 A^i}{\partial x^l \partial x^m} = \frac{4\pi}{c} j^i, \quad \text{with } A^i = 0, x^\alpha \notin D_\alpha \text{ and } A^i \not\equiv 0, x^\alpha \in D_\alpha. \qquad (12)$$

It was shown in [9] that there exist such 'source' functions $j^i$ for which the solutions $A^i$ satisfy to Eq.(10) with edge conditions. The em fields $A^i$ excited by stable OSC are *self-consistent* with their sources $j^i$ and *compact* in space in the continuum of the observer. Analysis of solutions of Eq.(12) shows that usually only discrete space sizes of compact fields are permitted to compensate em fields with some fixed frequencies outside space region $D_\alpha$. Edge conditions of Eq.(12) define the *quantization* of excited em fields, so stable OSC system has only discrete steady states. Distinct states are not stable, so, OSC system may pass from one stable state to another with exciting or absorbing em quanta. Figure 8 illustrates the process of excitation of the em fields in the continuum of the observer by the enclosed OSC object. There are also shown two quanta of em field excited during state changing and moving out from the stable OSC system.

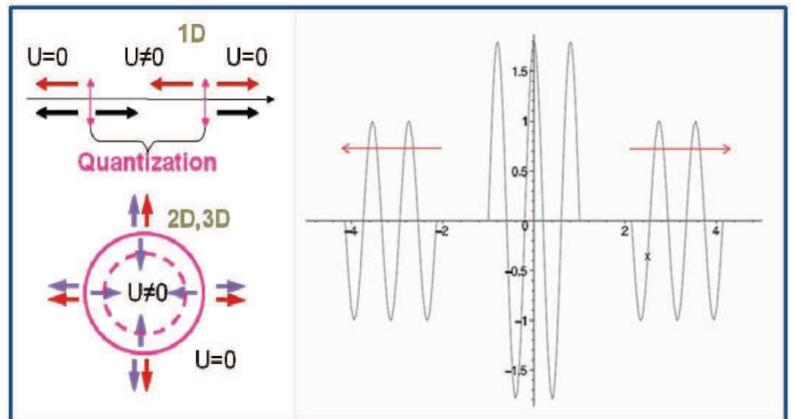

Figure 8. Excited em fields of OSC objects

---

[1] Excluding the first term from variation, one may lose many interesting solutions, but now we are investigating the principle question of existence at least some of steady states.

Similar self-consistent compact em fields exist, and one can even observe them in macro-objects. They exist in so-called optical *autosolitons* (N.N. Rosanov [10], B.S. Kerner [11]). Certainly, optical autosolitons exist in non-linear optical media and are quite different from the OSC by nature, but their excited em fields are also compact and self-consistent. They also possess the explicit quantum characteristics. Moreover, the internal structures of the autosolitons are quite similar to those of OSC physical objects shown on Fig.7. The *fundamental* autosolitons look like the stable state OSC (variant 1) and *vortical* ones are similar to closed chains of unstable states (variants 2,3…). The autosoliton needs the threshold energy for 'bearing', i.e. to come from 'open' chain (variant 0) to 'close' ones. It really looks like the model of physical vacuum and bearing of elementary particles. There are many other interesting coincidences of these so different physical objects. These coincidences are defined by similar self-consistent structure of excited fields of these objects. It is quite good sign in our analysis and may be considered as some indirect experimental confirmation. It was shown in [9] that the Klein-Gordon-Fock equation may be considered as a particular case of Eq.(12). The KGF equations have no steady-state solutions, but Eqs.(12) have.

The excited compact self-consistent em fields of the enclosed OSC possess the explicit quantum properties for the observer, which is a consequence of the energy conservation law in the system of the observer. These fields are good candidates for a role of *weak interactions*. Experiments confirm such interpretation by a fact of equivalence of weak and electromagnetic interactions at high energy levels. Thus, in OSC we already eliminate the second 'fundamental' interaction. Anyway, it was mentioned before that we had some doubts about the 'fundamentality' of strong and weak forces even without OSCs, because of their 'disappearing' with annihilation process.

As far as such interactions through the quantification conditions are connected with the spatial sizes of the OSC objects, so in macro-scale, scale of the galaxies and whole Universe, these em fields need to have some correspondences, for example, with the *relic radiation*. This 'relic' radiation differs by nature from registered streams of particles from 'empty' space regions of the Universe, which may be excited by OSC objects from non-observable regions excited. Unfortunately, analysis of Eq.(12) for cosmological objects may be more complicated, because, generally, it is not known whether the conditions of OSC stability and boundedness are fulfilled or not, because, at a moment to define these boundaries is complicated. This way, it seems quite interesting to identify the Large scale structures (F.S. Labini [12], P. Teericorpi [13]) and investigate them from OSC point of view.

# 6. Foundations of modern physics

In spite of all successes of both relativistic and quantum theories, these 'cornerstones' of modern physics can't be acknowledged as noncontradictory, self-consistent and systematic. Not only backgrounds of these theories contradict to each other, but also there are a lot of contradictions inside them [3,14,15]. For example, A.Einstein did not believe in the validity of quantum mechanics because of the EPR-like effects [16], but further experiments had confirmed the quantum mechanics conclusions. In spite of this, the Nobel prise winner M. Gell-Mann considered the quantum physics as *"an anti-intuitive discipline ... full of mysteries and paradoxes, which we do not understand quite well, but are able to use ... limits, in which, as we suppose, any correct theory should be included"* [17]. In one's turn, a lot of 'conceptual problems' are also in relativistic theories [3], which lead to the 'absurd Universe' of modern cosmology, violation of energy conservation law, etc. Even basic physical concepts of modern physics are contradictory. For example, we know a lot of 'elementary' particles and four 'fundamental' interactions, but 'elementary' particles can annihilate and, so, 'elementary' particles and two 'fundamental' interactions may simply transform to quanta of em field. In this connection, it is logical to ask what do notions 'elementary' and 'fundamental' mean? On the other hand, the invariance principle of special relativistic theory (SR) declined the 'absolute frame of references' in SR's physics. However this 'absolute frame of references' declined in SR is in fact introduced again in GR. Anyway, in SR the 'origin' of the inertiality is not understood. Therefore, it seems that conclusions contradict to backgrounds, which is a sign of inconsistency. The main Paradox is that both relativistic and quantum theories have found many excellent experimental confirmations. That is why modern physisists need to implement such contradictory physical reality description.

From this point of view, the main achievement of the OSC hypothesis is the possibility of elimination of contradictions between and inside relativistic and quantum approaches and even their unification, the possibility to introduce the noncontradictory description of the physical reality. Remarkable that it is possible without changing basic principles of relativistic and quantum approaches. Only extending to OSCs the relativistic principles one can achieve also their quantum physics description. It looks like only OSCs can make our physical Worldview systematic, can create a noncontradictory selfconsistent logical system of reality description. From OSC point of view it is clearly seen the limits of relativistic and quantum approaches.

Indeed, relativistic and quantum approaches are implemented inside the observer's continuum, but contradictions exist mainly in perception of different OSC objects. Extending the relativistic principle of Eq.(1) to the OSC system (Eq.(7)), one may easiely come to the quantum physics description (Eq.(6,10)). Quantization is the particularity of the OSC perception by the observer and is a consequence of an energy conservation law in the observer's continuum. Stable states are not postulated, but follow from the OSC methodology. Strong (Eq.(6)) and weak (Eqs.(10,10a)) interactions are not 'fundamental' in OSC: strong interactions are only the perception of the OSC topometry and weak interactions are simply excited compact self-consistent em fields. 'Elementary' particles are not 'elementary' at all, it is the same 'matter', but organized in different way, in different states. As far as the OSC objects are principally percieved as 'non-local', so there are no conceptual problems to explain their duality - 'wave' and 'particle' properties.

The 'relativistic' problems of the inertiality 'origin' and existence of 'absolute' frame of references are also easiely solved in OSC. The 'absolute' frame is defined by OSC objects and exists in every OSC, but it is not 'absolute' for another OSC. Hence, every observer has 'absolute' frame of references, but this frame does not coincide with one from different OSC. This way the 'inertiality' problem simply does not exist.

OSC gives a possibility to explain many conceptual problems in the standard cosmological model. Origins of dark matter and dark energy together with their correspondences with visual 'normal' matter clearly look through. The explanation of violations of conservation laws in standard models, for example, in expanding Universe, are available on frames of OSC by 'matter exchange' with OSC non-observable regions.

The OSC approach gives such powerful mathematical tool as theory of sets for the description of the physical reality. This theory gives many possibilities for physical reality description. As far as, generally, the ambiguous correspondences between physical objects from different OSCs are introduced, so chaotic, fractal, statistical methods of mathematical description may be effectively used inside OSC frames. Due to fundamental properties of OSC objects, the chaotic modeling and simulation are quite appropriate to the description of the OSCs. The mathematical analysis of the OSC with help of the theory of sets is universal enough and may include many mathematical approaches of modern physics. It was pointed out before, that, for example, additional hidden, compact spaces used in many modern physical theories may be considered as a particular case of the OSC.

Within the framework of OSC approach, it is possible to overcome frames of our observable 'material World', Universe, and to reach the realities inaccessible to us essentially, going out of the limits of observable physical reality to other ones. It may be considered as connecting of fields of investigation of fundamental science, philosophy and even religion. The religious paradigm of the logical wholeness of the material and non-material Worlds gives us hopes and perspectives for further investigations.

## Conclusions

It was found that the off-site continuums are available for the cosmological models of the galaxies and the Universe. Many conceptual problems of the standard cosmological model, such as 'dark matter' and 'dark energy', violations of energy-momentum conservation laws, may find physical explanation. On frames of OSC approach many correlations between relativistic and quantum physics and, hence, between cosmological and quantum physical objects were found. Such point of view may appear useful for the investigation of conceptual problems of modern cosmology.

Remarkable, that OSC approach can eliminate many contradictions in foundation of modern physics by unification of relativistic and quantum approaches without changing their backgrounds. From OSC point of view it is clearly seen the limits of relativistic and quantum approaches. It may be a way to non-contradictory description of physical reality. It looks like within the framework of OSC approach it is possible to overcome frames of our observable material World, Universe, and to reach the realities essentially inaccessible to us.

## References

1. Schrödinger E. Space-time structure / Cambridge at the University press, 1950.
2. Landau L.D., Lifshitz E.M. Theoretical Physics: The Classical Theory of Fields, vol.2 (Course of Theoretical Physics Series) / Pergamon Press, 1988.
3. Baryshev Y.V. Conceptual Problems of the Standard Cosmological Model / astro-ph/0509800v1, 2005.
4. Chernin A.D., et al. Local dark energy: HST evidence from the vicinity of the M81/82 galaxy group / Astrophysics, vol.50,No.4,2007.
5. Chernin A.D., et al. Local dark energy: HST evidence from the expansion flow around Cen A/M83 galaxy group / astro-ph/0704.2753, 2007.
6. Soloviev L.D. Straight-line string with the curvature / hep-th/9503156, 1995.
7. Soloviev L.D. Relativistic quantum model of confinement and the current quark masses / hep-th/9803483, 1998.
8. Polyakov A. Fine Structure of Strings / Nucl.Phys. B268, 406, 1986.
9. Novikov-Borodin A.V. Off-site continuums and methods of their mathematical description / physics.gen-ph/07060085v1, 1 June 2007.
10. Rosanov N.N. The World of Laser Solitons. – Nature,6, 2007.
11. Kerner B.S., Osipov V.V. Autosolitons: a new approach to problems of self-organization and turbulence. – Springer, 1994.
12. Labini F.S. Correlation and Clustering in the Universe. – Proc.of the Int.Conf."Problems of Practical Cosmology" (http://ppc08.astro.spbu.ru), Saint-Petersburg, Russia, June 23-27, 2008.
13. Teerikorpi P. A Brief History of Large Scale Structures: from the 2D Sky to the 3D Maps. – Proc.of the Int.Conf."Problems of Practical Cosmology" (http://ppc08.astro.spbu.ru), Saint-Petersburg, Russia, June 23-27, 2008
14. Butterfield J. The State of Physics: 'Halfway through the Woods' / Journal of Soft Comp., IQS Associat., 1999.
15. Meskov V.S. Syntactical analysis of the theory completeness problem: the quantum mechanics and arithmetic / in: Investigations on non-classical logic, VI Soviet-Finnish colloquium, Nauka, Moscow, 1989.
16. Einstein A., Podolsky B., Rosen N. Can Quantum-Mechanical Description of Physical Reality be considered Complete? / Physical Review 47, 1935, pp. 777.
17. Gell-Mann M. Questions for the Future / in: The Nature of Matter, Wolfson College, Oxford 1981.